\documentstyle[12pt,aaspp4]{article}
\begin{document}

\title{MODELLING THE GALACTIC BAR USING RED CLUMP GIANTS}

\author{K.~Z.~Stanek\altaffilmark{1}}
\affil{Princeton University Observatory, Princeton, NJ 08544--1001}
\affil{\tt e-mail: stanek@astro.princeton.edu}
\altaffiltext{1}{On leave from N.~Copernicus Astronomical Center, 
Bartycka 18, Warszawa 00--716, Poland}
\author{A. Udalski, M. Szyma\'nski, J. Ka\l u\.zny and M. Kubiak}
\affil{Warsaw University Observatory
Al. Ujazdowskie 4, 00--478 Warszawa, Poland}
\author{M. Mateo}
\affil{Department of Astronomy, University of Michigan,
821 Dennison  Bldg., Ann Arbor, MI~48109--1090}
\author{and}
\author{W.~Krzemi\'nski}
\affil{Carnegie Observatories, Las Campanas Observatory,
Casilla 601, La Serena, Chile}

\begin{abstract}

The color-magnitude diagrams of $\sim 7 \times 10^5$ stars obtained for 12
fields across the Galactic bulge with the OGLE project reveal a well-defined
population of bulge red clump giants.  We find that the distributions of the
apparent magnitudes of the red clump stars are systematically fainter when
moving towards lower galactic $l$ fields. The most plausible explanation of
this distinct trend is that the Galactic bulge is a bar, whose nearest end lies
at positive galactic longitude.  We model this Galactic bar by fitting for all
fields the observed luminosity functions in the red clump region of the
color-magnitude diagram.  We find that almost regardless of the analytical
function used to describe the 3-D stars distribution of the Galactic bar, the
resulting models have the major axis inclined to the line of sight by
$20-30\deg$, with axis ratios corresponding to
$x_0\!:\!y_0\!:\!z_0=3.5\!:\!1.5\!:\!1$. This puts a strong constraint on the
possible range of the Galactic bar models.  Gravitational microlensing can
provide us with additional constrains on the structure of the Galactic bar.

\end{abstract}


\section{INTRODUCTION}

There is a mounting evidence that the Galactic bulge is a triaxial structure,
or a bar. This was first postulated by de Vaucouleurs~(1964), based on
similarities between the kinematics of the gas observed towards the Galactic
center and in other barred galaxies. However, the hypothesis that our Galaxy is
a barred galaxy was for a long time overshadowed by ``$3\; kpc$ expanding arm''
hypothesis, despite a number of papers arguing for bar's presence (e.g. Peters
1975; Liszt \& Burton 1980; Gerhard \& Vietri 1986). Only recently has the view
of our Galaxy as a barred spiral gained momentum, mostly due to the work of
Blitz \& Spergel (1991). They analyzed $2.4\;\mu m$ observations of the
Galactic center and showed convincingly that the observed asymmetry in the
galactic longitude distribution of surface brightness is naturally explained by
the bar with the near side in the first Galactic quadrant. Binney et al.~(1991)
have constructed a dynamical model for gas in the inner Galaxy, and their
resulting bar has the same orientation as that suggested by Blitz \&
Spergel~(1991) in the sense that the closer part of the bar is at positive
galactic longitudes. COBE-DIRBE multiwavelength observations of the Galactic
center (Weiland et al.~1994) confirmed the existence of the longitudinal
asymmetry discussed by Blitz \& Spergel~(1991). This data was used by Dwek et
al.~(1995) to constrain a number of analytical bar models existing in the
literature.
 
Star counts have also shown evidence for triaxial structure in the center of
the Galaxy. Nakada et al.~(1991) analyzed the distribution of IRAS Galactic
bulge stars and found asymmetry in the same sense as Blitz \&
Spergel~(1991). Whitelock \& Catchpole~(1992) analyzed the number distribution
of Mira variables in the bulge as a function of distance modulus and found that
the half of the bulge which is at positive galactic longitude is closer to us
than the other half. The observed stellar distribution could be modelled with a
bar inclined at roughly $45\deg$ to the line of sight.  Weinberg~(1992) used
AGB stars as star tracers and mapped the Galaxy inside the solar circle. He
found evidence for a large stellar bar with semimajor axis of $\approx\;5\;kpc$
and inclination placing the nearer side of the bar at positive galactic
longitudes.

A third thread of evidence comes from the gravitational microlensing towards
the Galactic bulge (Udalski et al.~1994; Alcock et al.~1995; Alard et
al. 1995).  Observed high microlensing rate can be accounted for by stars
placed in the near part of the Galactic bar microlensing the stars from the far
side of the bar (Kiraga \& Paczy\'nski 1994; Paczy\'nski et al.~1994b; Zhao,
Spergel \& Rich 1995).  There are many implications if this scenario is
correct, discussed latter in this paper. For recent reviews on the Galactic
bar see Gerhard (1996) and Kuijken (1996).

In this paper we use a color-magnitude data obtained by the Optical
Gravitational Lensing Experiment collaboration for 12 fields scattered across
the galactic bulge to construct the three-dimensional model of the mass
distribution in the Galactic bar.  The Optical Gravitational Lensing Experiment
(OGLE, Udalski et al.~1993; 1994) is an extensive photometric search for the
rare cases of gravitational microlensing of Galactic bulge stars by foreground
stars, brown dwarfs and planets. It provides a huge database (Szyma\'nski \&
Udalski 1993), from which color-magnitude diagrams have been compiled (Udalski
et al.~1993).  Stanek et al.~(1994; 1996) used the well-defined population of
bulge red clump stars to investigate the presence of the bar in our Galaxy.
Comparing extinction-adjusted apparent magnitudes of the red clump giants
observed at fields lying at $l=\pm5\deg$, we found that the red clump stars
lying at the positive Galactic longitudes are systematically brighter by
$\sim0.4\;mag$ than the stars lying at negative $l$.  A bar-shaped bulge whose
nearer side is at the positive Galactic longitude most easily explains this
offset. This agrees with Blitz \& Spergel~(1991) and other works.

The paper is organized as follows. In Section 2 we discuss the basis for using
the red clump stars as a distance indicator.  In Section 3 we discuss the data
used for the bar modelling.  In Section 4 we discuss various analytical models
of the bar.  In Section 5 we constrain these models using our data. In
Section 6 we discuss various factors which might affect the fits. In Section 7
we discuss some astrophysical implications of the constructed models, among the
others for the gravitational microlensing optical depth.

\section{RED CLUMP GIANTS AS A DISTANCE INDICATOR}

Before we model the Galactic bar using red clump stars, we shall discuss the
evidence that the red clump stars would be a good distance indicator for the
Galactic bulge population.  Red clump stars are the equivalent of the
horizontal branch stars for a metal rich population, i.e. relatively low mass
stars burning helium in their cores.  From stellar evolution theory (e.g.
Seidel, Demarque \& Weinberg 1987; Castellani, Chieffi \& Straniero 1992) we
expect the red clump stars to have a narrow luminosity distribution with weak
dependence on age and the metallicity.  For given metallicity and helium
abundance, the total difference in mean clump luminosity for ages between 1 and
$10\;Gyr$ is $\sim 0.1\;mag$ (Seidel et al.~1987). The effect of metallicity on
red clump giants luminosity and color is also relatively weak (see Sarajedini,
Lee \& Lee 1995, their Figs.4,5). Therefore, red clump stars form a suitable
population with which to investigate the properties of high-metallicity
systems, like the Galactic bulge.

\begin{figure}[t]
\plotfiddle{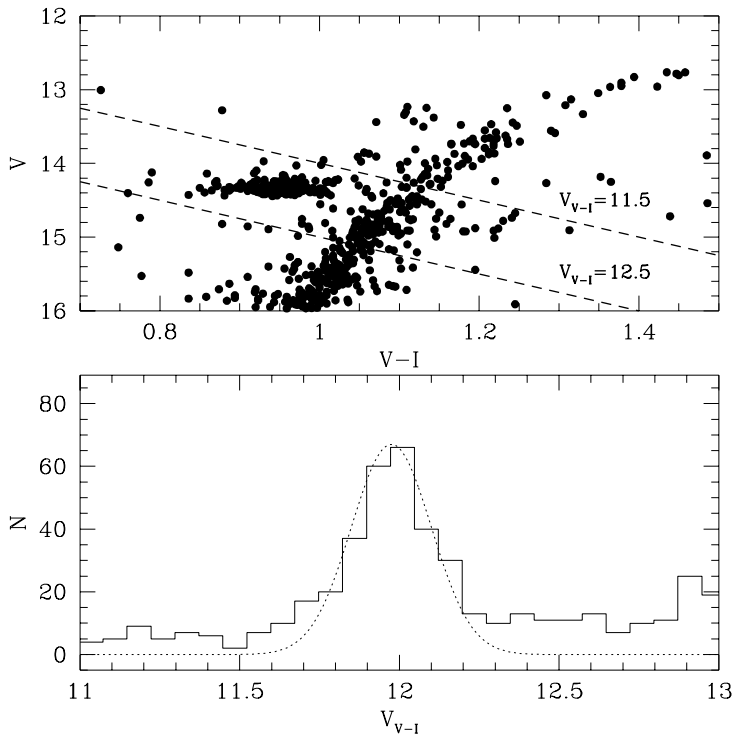}{8.5cm}{0}{50}{50}{-160}{-75}
\caption{Red clump dominated part of the CMD for the globular cluster
47 Tuc (upper panel). The two dashed lines correspond to the values of the
extinction-adjusted parameter $V_{_{V-I}}$ of 11.5 and 12.5.  In lower panel we
show histogram of $V_{_{V-I}}$ values in the vicinity of the red clump. Dashed
line corresponds to a Gaussian with $\sigma=0.12\;mag$.}
\label{fig1}
\end{figure}

\begin{figure}[t]
\plotfiddle{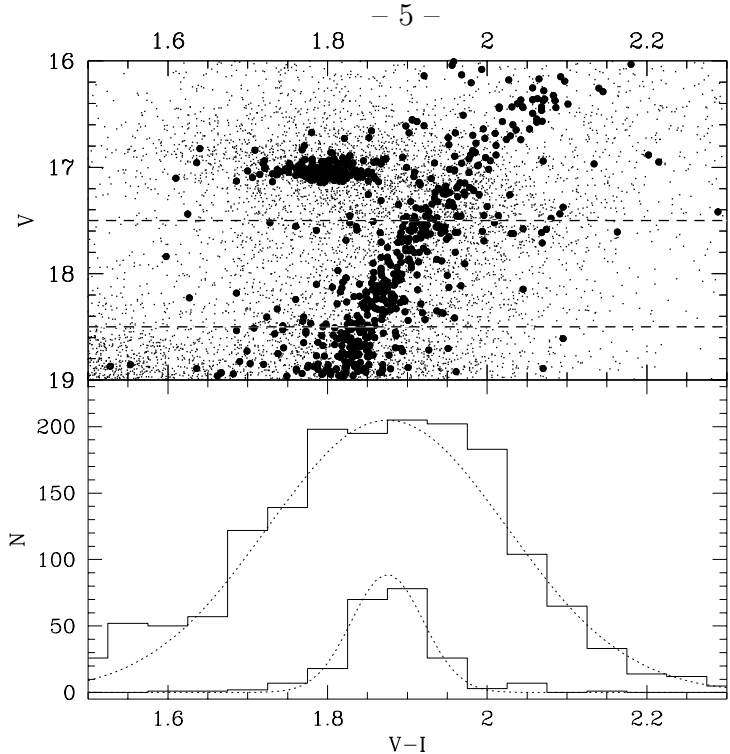}{8.5cm}{0}{50}{50}{-160}{-75}
\caption{Comparison of the width of the red giant branch between the
globular cluster 47 Tuc and one of the OGLE fields BW4. In upper panel we show
the red clump dominated parts of both CMDs. The 47 Tuc CMD (big dots) was
shifted so as to match the red clump position of the BW4 CMD (small dots). In
lower panel we show histogram of $V-I$ colors for both red giant branches
(limited in $V$ by two dashed lines shown in the upper panel).  The red giant
branch of 47 Tuc is narrow, $\sigma\approx0.05\;mag$, and it is much broader
for the BW4 field, with $\sigma\approx0.15\;mag$.}
\label{fig2}
\end{figure}

Because of high and variable interstellar extinction towards the Galactic
bulge, we use extinction-adjusted parameter $V_{_{V-I}}$ in our analysis (see
Eq.1). We would like to be able to say something about the intrinsic width of
the red clump population. For that, we use $V,V-I$ color-magnitude diagrams
(CMDs) of the globular cluster 47 Tuc obtained by one of the authors (JK) using
2.5-meter DuPont telescope at the Las Campanas Observatory.  These CMDs are
described in more detail by Ka\l u\.zny \& Wysocka (1996).  47 Tuc is a metal
rich globular cluster with $\rm [Fe/H]\approx-0.7$ (Zinn \& West 1984) and age
of $13.0\;Gyr$ (Sarajedini et al.~1995).

We first construct a histogram of $V_{_{V-I}}$ values for stars in the red
clump dominated region of the 47 Tuc CMD (Fig.\ref{fig1}).  The definition of the
$V_{_{V-I}}$ parameter causes that, along with the red clumps giants we are
interested in, also stars from the red giant branch which happen to have the
same values of $V_{_{V-I}}$ are included into histogram. The resulting peak in
the luminosity function is quite narrow --- for comparison we plotted a
Gaussian with $\sigma=0.12\;mag$.

We want also to be able to estimate what are the effects of possible
metallicity spread (McWilliam \& Rich 1994; Minniti et al.~1995) on the red
clump width. To do so, we selected one of Baade's Window fields observed by
OGLE, BW4, which has relatively uniform extinction (see Stanek 1996). We than
shifted the CMD of 47 Tuc to match the location of the red clumps giants in the
BW4. This is shown in the upper part of Fig.\ref{fig2}.  We then plotted the
histogram of $V-I$ color for red giant branch stars from both the cluster and
the BW4 field (Fig.\ref{fig2}, lower part). The red giant branch of 47 Tuc is
very narrow, $\sigma\approx0.05\;mag$, and it is much broader for the BW4
field, with $\sigma\approx0.15\;mag$. This combined with the $\sigma=0.12\;mag$
measured in the previous paragraph gives us the intrinsic width of the red
clump giants population in the Galactic bulge
$\sigma_{RC}=\sqrt{0.12^2+0.15^2}\approx 0.2\;mag$. We will use this number
later in this paper.

\section{THE DATA}

\begin{figure}[t]
\plotfiddle{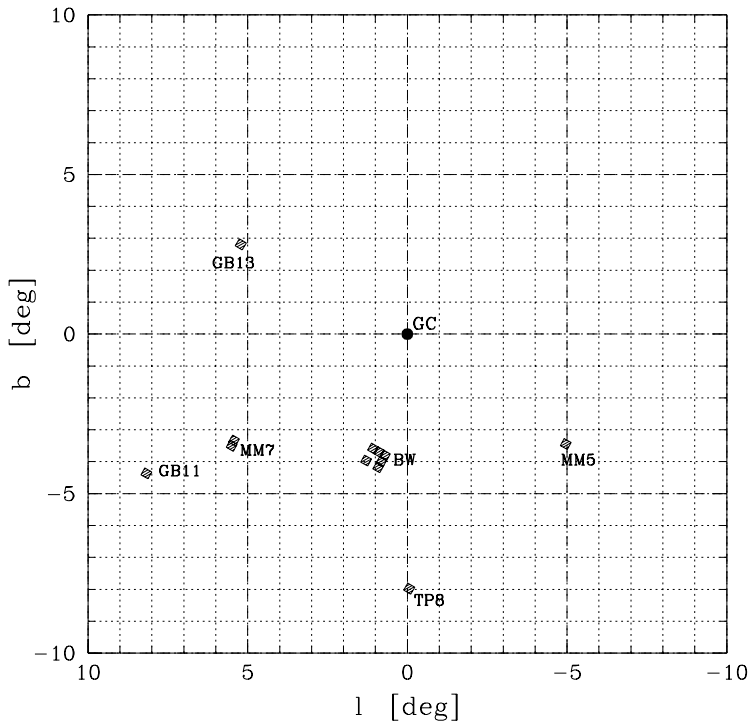}{8.5cm}{0}{50}{50}{-160}{-75}
\caption{Schematic map of the 12 fields used in this paper to model the
Galactic bar. The grid shown corresponds to the Galactic coordinates.
GC denotes the Galactic Center, BW -- Baade's Window. For coordinates
of the fields see Table~1.}
\label{fig3}
\end{figure}

We use a color-magnitude data obtained by the Optical Gravitational Lensing
Experiment collaboration for 12 fields scattered across the galactic bulge to
construct the 3-D model of the mass distribution in the Galactic bar. The
schematic map of the fields' location in the Galactic coordinates is shown in
Fig.\ref{fig3} and the coordinates of the fields are given in Table~1. We use 10
color-magnitude diagrams (CMDs) from Udalski et al.~(1993) (fields BW1-6,
MM5-B, MM7-AB, TP8) and two CMDs obtained last year specifically for the
current project (fields GB11, GB13). All the data were obtained using the
1-meter Swope telescope at the Las Campanas Observatory, operated by the
Carnegie Institution of Washington, and $2048\times 2048$ Ford/Loral CCD
detector with the pixel size 0.44 arcsec covering $15' \times 15'$ field of
view.  We did not use fields BW7 and BWC of Udalski et al.~(1993) because they
each contain a globular cluster, which could potentially contaminate our star
counts. We also excluded fields BW8 and MM5-A because of very strong CCD
bleeding columns caused by bright stars.  Most of each CMD is dominated by
bulge stars, with a distinct red clump, red giant, and turn-off point stars.
The part of the diagram dominated by disk stars for nine BW fields was analyzed
by Paczy\'nski et al.~(1994a). Stanek et al.~(1994) used well-defined
population of bulge red clump giants in nine BW fields and four MM fields to
find an evidence for the Galactic bar. This paper is an extension of this
earlier work.

\begin{planotable}{lrrrrr}
\tablewidth{30pc}
\tablecaption{Observed Fields}
\tablehead{ \colhead{Field} & \colhead{$\alpha_{2000.0}$} &
\colhead{$\delta_{2000.0}$}
& \colhead{$l$}  & \colhead{$b$} & \colhead{No. of} \\
\colhead{} &  \colhead{} &  \colhead{} & 
\colhead{} & \colhead{} & \colhead{RC stars} }
\startdata
BW1    & 18:02:19.9 & $-$29:49:28 & 1.0764    & $-$3.5894 & 5,496 \nl
BW2    & 18:02:20.6 & $-$30:15:18 & 0.7017    & $-$3.8028 & 4,380 \nl
BW3    & 18:04:19.5 & $-$30:15:22 & 0.9103    & $-$4.1769 & 3,865 \nl
BW4    & 18:04:18.8 & $-$29:49:12 & 1.2908    & $-$3.9625 & 4,549 \nl
BW5    & 18:02:20.1 & $-$30:02:27 & 0.8878    & $-$3.6961 & 5,382 \nl
BW6    & 18:03:20.8 & $-$30:15:10 & 0.8099    & $-$3.9909 & 5,045 \nl
MM5--B & 17:47:24.3 & $-$34:57:38 & $-$4.9577 & $-$3.4479 & 3,844 \nl
MM7--A & 18:10:52.1 & $-$25:54:16 & 5.4272    & $-$3.3385 & 3,572 \nl
MM7--B & 18:11:39.4 & $-$25:55:08 & 5.4996    & $-$3.5008 & 3,493 \nl
TP8    & 18:17:55.0 & $-$32:54:03 & $-$0.0583 & $-$7.9773 &   855 \nl
GB11   & 18:20:34.6 & $-$23:58:45 & 8.1703    & $-$4.3672 & 2,339 \nl
GB13   & 17:46:58.5 & $-$23:01:14 & 5.2157    & 2.8113    & 5,144
\enddata
\tablecomments{Number of red clump stars was taken as number of stars
fulfilling the inequalities (\ref{eq:select}), or (\ref{eq:select2}) for the
field TP8 and (\ref{eq:select1}) for the field GB13.}
\end{planotable}

\begin{figure}[t]
\plotfiddle{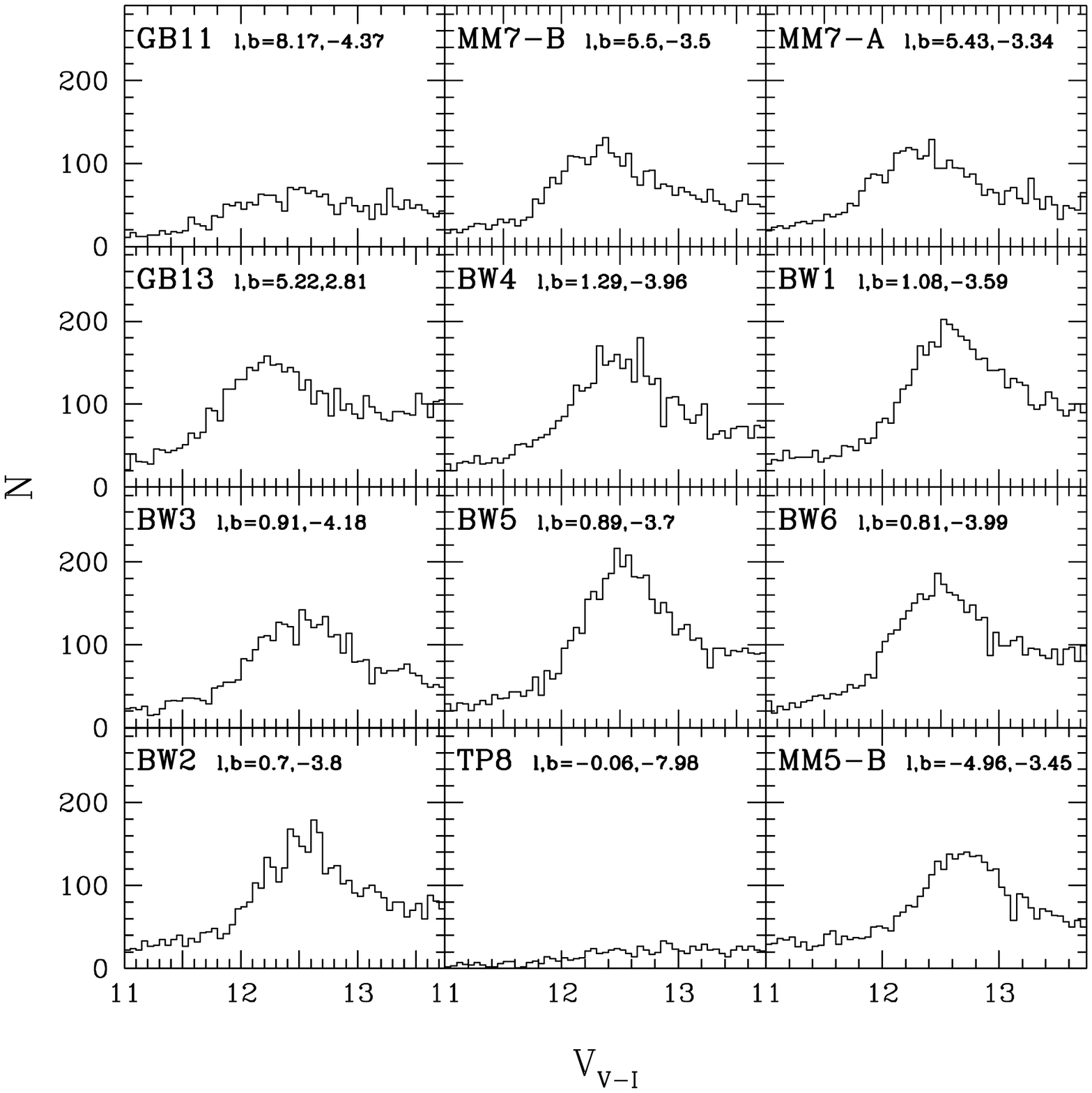}{8.5cm}{0}{50}{50}{-160}{-75}
\caption{Histograms of $V_{_{V-I}}$ values in the red clump region
(Eq.2,3,4) for all 12 fields, ordered by decreasing Galactic longitude.  Please
note the systematic shift, towards the larger values of $V_{_{V-I}}$, between
the red clump giants distribution as the $l$ decreases -- this is the bar
signature already seen by Stanek et al.~(1994; 1996).}
\label{fig4}
\end{figure}

To analyze the distribution of bulge red clump stars in a quantitative
manner, we use the extinction-insensitive $V_{_{V-I}}$ parameter
(Paczy\'nski et al.~1994a; Stanek et al.~1994)
\begin{equation}
  V_{_{V-I}} \equiv V - 2.5 ~ (V-I),
\label{eq:free}
\end{equation}
where we use reddening law $ E_{_{V-I}} = A_{_V}/2.5 $, following
Wo\'zniak \& Stanek (1996) and Stanek (1996). The parameter $V_{_{V-I}}$
has been defined so that if $A_{_V}/E_{_{V-I}}$ is independent of location
then for any particular star its value is not affected by the unknown
extinction (Wo\'zniak \& Stanek 1996). Then we consider only the
region of the CMD clearly dominated by the bulge red clump stars:
\begin{equation}
1.5  < V-I < 2.4 ~~;~~ 11.0 < V_{_{V-I}} < 13.75 
\label{eq:select}
\end{equation}
For one of the fields, TP8, the average interstellar extinction is
clearly lower than for the other fields, so for TP8 the selected
region is:
\begin{equation}
1.1  < V-I < 2.0 ~~;~~ 11.0 < V_{_{V-I}} < 13.75,
\label{eq:select2}
\end{equation}
on the other hand, for one of the new fields, GB13, the average interstellar
extinction is clearly higher than for the other 11 fields, so for GB13 the
selected regions is:
\begin{equation}
1.9  < V-I < 2.8 ~~;~~ 11.0 < V_{_{V-I}} < 13.75 
\label{eq:select1}
\end{equation}
Please note that in Eqs.(\ref{eq:select}--\ref{eq:select1}) the range of
$V_{_{V-I}}$ is always the same.  If the value of the coefficient of the
selective extinction is indeed constant throughout the Galactic bulge, as found
by Wo\'zniak \& Stanek (1996), then these different selections for TP8 and GB13
fields should not affect our calculations.  All stars that satisfy the
inequalities (\ref{eq:select}) and (\ref{eq:select2},\ref{eq:select1}) were
counted in bins of $\Delta V_{_{V-I}} = 0.05 $. The result appears in Fig.\ref{fig4},
where we see the number of stars as a function of $V_{_{V-I}}$ for all the
fields, ordered by decreasing Galactic longitude.  Please note the systematic
shift between the red clump giants distribution as the $l$ decreases, as
already seen by Stanek et al.~(1994; 1996).  Please also note that the
$V_{_{V-I}}$ histogram of the field GB13 is not shifted comparing to the MM7
fields, which indicates that the coefficient of the selective extinction is
indeed constant throughout the Galactic bulge.

\section{TRIAXIAL BAR MODELS}

We seek to reproduce the observed luminosity function in each of the 12
fields (Fig.\ref{fig4}) assuming some analytical function for the Galactic bar
density distribution and fitting some simple form of the intrinsic luminosity
function, the same in all the fields, for the stars in the red clump region.

For the Galactic bar density distributions we follow Dwek et al.~(1995)
approach and select three families of the analytical functions
(for detailed description see Dwek et al.):

\begin{enumerate}
\item {\em Gaussian type functions} (G), described by
\begin{eqnarray}
\rho_{G1}(x,y,z) & = & \rho_0 \exp(-r^2/2)
\label{eq:gauss} \\
\rho_{G2}(x,y,z) & = & \rho_0 \exp(-r_s^2/2) \\
\rho_{G3}(x,y,z) & = & \rho_0 r^{-1.8}\exp(-r^3)
\end{eqnarray}

\item {\em exponential-type functions} (E), described by
\begin{eqnarray}
\rho_{E1}(x,y,z) & = & \rho_0 \exp(-r_e) \\
\rho_{E2}(x,y,z) & = & \rho_0 \exp(-r) \\
\rho_{E3}(x,y,z) & = & \rho_0 K_0(r_s)
\label{eq:exp}
\end{eqnarray}

\item {\em power-law type functions} (P), described by
\begin{eqnarray}
\rho_{P1}(x,y,z) & = & \rho_0 \left(\frac{1}{1+r}\right)^4 \\
\rho_{P2}(x,y,z) & = & \rho_0 \frac{1}{r(1+r)^3} \\
\rho_{P3}(x,y,z) & = & \rho_0 \left(\frac{1}{1+r^2}\right)^2
\label{eq:power}
\end{eqnarray}

\end{enumerate}

where:

\begin{eqnarray}
r & \equiv &
\left[\left(\frac{x}{x_0}\right)^2+\left(\frac{y}
{y_0}\right)^2+\left(\frac{z}{z_0}\right)^2\right]^{1/2} \\
r_e & \equiv &
\left(\frac{\left|x\right|}{x_0}+\frac{\left|y\right|}
{y_0}+\frac{\left|z\right|}{z_0}\right) \\
r_s & \equiv &
\left\{\left[\left(\frac{x}{x_0}\right)^2+\left(\frac{y}{y_0}\right)^2
\right]^2+\left(\frac{z}{z_0}\right)^4\right\}^{1/4}
\end{eqnarray}
All these functions are allowed to be rotated in the plane of the Galaxy by the
angle $\alpha$, related to Dwek's et al. angle $\alpha_D$ by
$\alpha=90-\alpha_D$ ($\alpha=0\deg$ corresponds to the major axis of the bar
pointing at us). We also allow for the tilt out of the Galactic
plane $\beta$, corresponding to Dwek's et al. angle $\beta$.  We assume the
distance to the center of the Galaxy $8.0\;kpc$.

We parametrize the fitted luminosity function in the form
\begin{equation}
\Phi(L)=\left(\frac{N_0}{L_{\odot}}\right)\left(\frac{L}{L_{\odot}}
\right)^{-\gamma}+\frac{N_{RC}}{\sigma_{RC}\sqrt{2\pi}}\exp\left[-
\frac{(L-L_{RC})^2}{2\sigma_{RC}^2}\right]~\;
\left[L^{-1}_{\odot}\right].
\label{eq:lumfun}
\end{equation}
The power-law with index $\gamma$ is intended to represent the underlying broad
population of stars, and the Gaussian is intended to represent red clump stars.
For a given $V_{_{V-I}}$ bin, the number of stars in that bin is then given by
\begin{equation}
N(V_{_{V-I}})=C_1\int_{D_{MIN}}^{D_{MAX}}\rho(D_s)D_s^2\Phi(L)L\;dD_s,
\label{eq:integ}
\end{equation}
where we take $D_{MIN}=3\;kpc,~D_{MAX}=13\;kpc,~C_1$ is a constant, $D_s$ is a
distance from the observer to the source, and $L$ is given by
\begin{equation}
L=C_2 D_s^2\; 10^{-0.4 V_{_{V-I}}},
\end{equation}
$C_2$ being another constant. In equation~(\ref{eq:integ}) we assume 
that the number of observable stars is everywhere proportional to the 
density of matter and that the intrinsic luminosity function is independent
of location, which gives a constant $(M/L)$ throughout the bar.

We now have a model of the mass distribution with four or five free parameters
(three axis plus one or two angles, Eqs.\ref{eq:gauss}-\ref{eq:power}) and the
intrinsic luminosity function with four free parameters (two for the underlying
power law plus two for the Gaussian red clump, Eq.\ref{eq:lumfun} -- the
intrinsic width of the Gaussian is not fitted, see below), total of 8 or 9 free
parameters.  We do not fit the intrinsic width of the red clump distribution
$\sigma_{RC}$ -- we pre-set it to three values corresponding to $0.1, 0.2$ and
$0.3\;mag$, with $0.2\;mag$ being the preferred value based on analysis in
Section~2. We should mention here that the luminosity function from
Eq.\ref{eq:lumfun} is only approximately Gaussian when transferred into
magnitudes. On the other hand we have 12 fields with 55 bins in each of the
fields, a total of 660 bins. This gives us number of degrees of freedom
$N_{DOF}=660-8=652$, which should be adequate to fit the model of the Galactic
bar. Please note that, unlike the case of COBE-DIRBE data, we have the
information about all three dimensions of the bar, but we lack COBE-DIRBE
spatial coverage.

\section{BAR MORPHOLOGY}

Table~2 presents some of the fitted parameters in the case with the value of
the assumed intrinsic width of the red clump distribution
$\sigma_{RC}=0.2\;mag$.  As we mentioned above, we also investigated cases with
$\sigma_{RC}=0.1,0.3\;mag$ and we will discuss some aspects of this below.  In
all three cases we have two sub-cases: one when we allow only eight free
parameters (with $\beta=0$) and one in which we allow for the tilt out of the
Galactic plane ($\beta\neq0$).

\begin{planotable}{ccrrrrc}				       
\tablewidth{30pc}					       
\tablecaption{Parameters of the Best-Fit Bar Models for
$\sigma_{RC}=0.2\;mag$}	 
\tablehead{ \colhead{Model}   & \colhead{$\alpha$}   &	       
\colhead{$\beta$} & \colhead{$x_0$}      & 		       
\colhead{$y_0$}   & \colhead{$z_0$}      &		       
 \colhead{$\chi^2/N_{DOF}$} \\   
\colhead{}        & \colhead{[$\,\deg\,$]}     &	       
\colhead{[$\,\deg\,$]}  & \colhead{[$\,pc\,$]}      &		 
\colhead{[$\,pc\,$]}   & \colhead{[$\,pc\,$]}  & \colhead{} } 
\startdata							 
G1  & 24.6 & \nodata & 1514 &  691 &  463 & 2.60  \nl
    & 24.5 & 0.5     & 1529 &  676 &  460 & 2.60  \nl
G2  & 24.9 & \nodata & 1239 &  609 &  438 & 2.90  \nl
    & 23.9 & 2.4     & 1293 &  565 &  423 & 2.84  \nl
G3  & 24.8 & \nodata & 3851 & 1698 & 1114 & 2.43  \nl
    & 24.7 & $-$0.1  & 3857 & 1705 & 1115 & 2.43  \nl
E1  & 20.7 & \nodata & 1665 &  691 &  297 & 2.47  \nl
    & 20.1 & $-$0.1  & 1680 &  704 &  296 & 2.47  \nl
E2  & 23.8 & \nodata &  899 &  386 &  250 & 2.37  \nl
    & 23.8 & 0.0     &  897 &  387 &  250 & 2.37  \nl
E3  & 23.9 & \nodata &  854 &  381 &  278 & 2.59  \nl
    & 24.0 & 0.4     &  857 &  376 &  276 & 2.59  \nl
P1  & 22.7 & \nodata &  724 &  299 &  189 & 2.27  \nl
    & 22.6 & $-$0.2  &  738 &  308 &  194 & 2.27  \nl
P2  & 22.6 & \nodata & 1056 &  435 &  275 & 2.27  \nl
    & 22.4 & $-$0.2  & 1061 &  443 &  278 & 2.27  \nl
P3  & 22.8 & \nodata & 1400 &  580 &  368 & 2.27  \nl
    & 22.7 & $-$0.2  & 1389 &  582 &  368 & 2.27  \nl
\enddata
\tablecomments{For every model two cases are shown: first with 8 free
parameters and second with $\beta\ne0$ (9 free parameters).}
\end{planotable}

We use $\chi^2$ statistics as both a fitting tool, and also as a relative
measure of the quality of each of the models. As we discuss later in this
paper, our best fit models have values of the $\chi^2/N_{DOF}\approx2.3$, which
indicates that $\chi^2$ should not be used as a measure of absolute goodness of
the fit. We therefore do not give formal errors of the fitted parameters in
Table~2 based on usual $\chi^2$ confidence levels.

\subsection{Comparison between the models}

Before we start discussing triaxial bar models, we should mention that we also
tried to fit axially-symmetric set of models, described by the same functions
as the triaxial bars, but with $x_0$ set to be the same as $y_0$. Not
surprisingly, axially symmetric models fitted our data substantially worse than
triaxial models, with the best fit model achieving $\chi^2/N_{DOF}\approx3.1$
compared to $\chi^2/N_{DOF}\approx2.3$ for the best triaxial models.

From Table~2 we can immediately see that none of the models fits the data
perfectly, i.e. $\chi^2/N_{DOF}$ is relatively high. We discuss possible
reasons for that in Section~6. Still, some analytical functions fit better then
others. We find that the whole class of power-law models (P1, P2, P3) gives us
the best fits to the data, for all three values of the $\sigma_{RC}$
investigated.  However, the values of $\chi^2/N_{DOF}$ for other analytical
functions, especially for the G3 and E2 models, are not much worse then for the
power-law models. What is somewhat surprising is that the G2 model is our
``worst fit'' model, given that this was Dwek et al. best fit model. We will
discuss agreement of our work with the analysis of COBE-DIRBE data in
Section~7.

\begin{figure}[t]
\plotfiddle{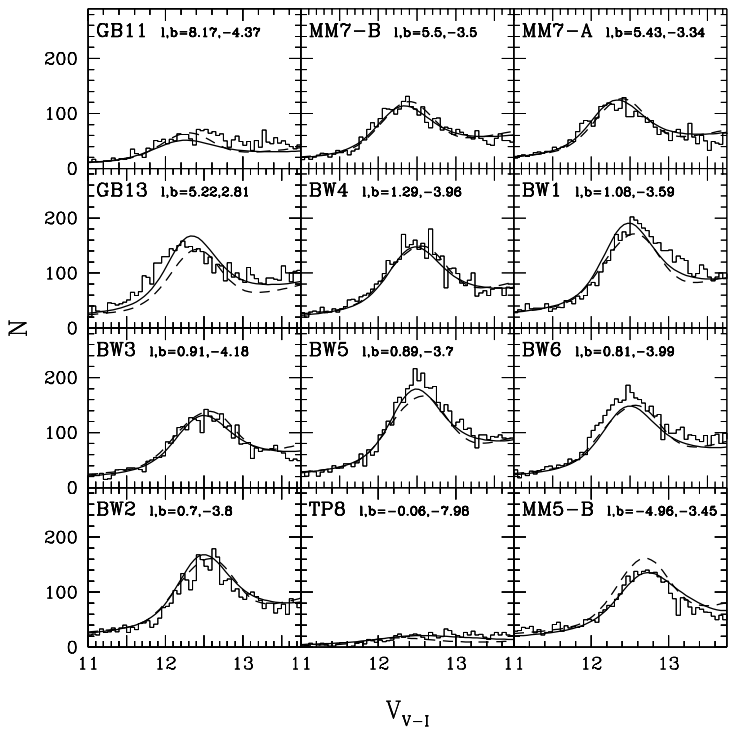}{8.5cm}{0}{50}{50}{-160}{-75}
\caption{The same histograms as in Fig.4, but with the best (P1 --
continues line) and worst (G2 -- dashed line) model fits shown. The assumed
value of the intrinsic width of the red clump for the fits is
$\sigma_{RC}=0.2\;mag$.}
\label{fig5}
\end{figure}

To illustrate how significant the differences are between the best and the
worst fit models, in Fig.\ref{fig5} we show the same histograms as in
Fig.\ref{fig4}, but with the best (P1 -- continues line) and worst (G2 --
dashed line) model fits shown. For some of the fields, like GB13 or MM5-B,
these differences are easily noticeable.

\begin{figure}[t]
\plotfiddle{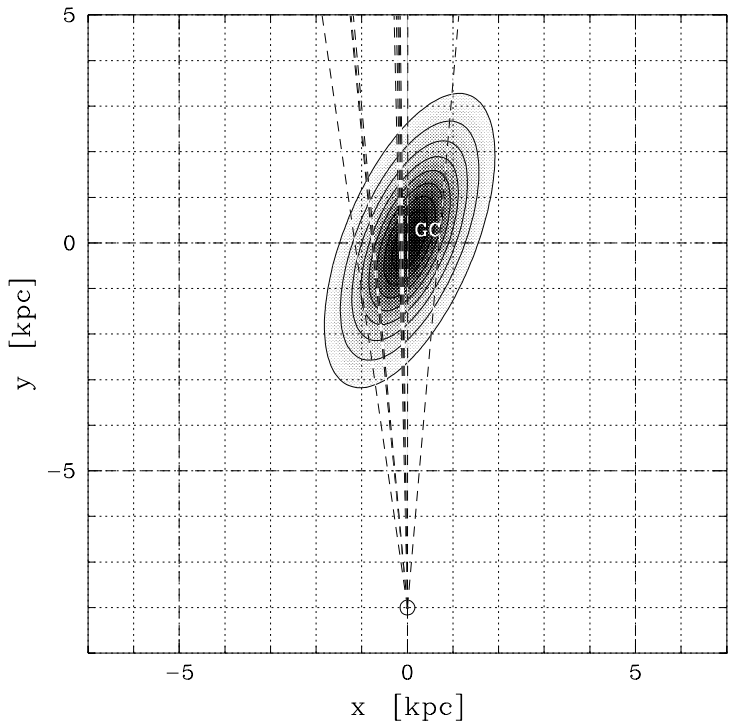}{8.5cm}{0}{50}{50}{-160}{-75}
\caption{Cut through the ``recommended''  E2 model at $700\;pc$ below the
Galactic plane, i.e. at the galactic latitude $b=-5\deg$ at the distance of
$8\;kpc$. With dashed lines we show the lines of sight for the 12 fields used
in this paper. Contours start at 10\% of the central value and increase by
10\%.}
\label{fig6}
\end{figure}

In Fig.\ref{fig6} we show a cut through the E2 (our ``recommended'' model of the
Galactic bar, see Section~7.1) model at $700\;pc$ below the Galactic plane,
i.e. at the galactic latitude $b=-5\deg$ at the distance of $8\;kpc$. With
dashed lines we show the lines of sight for the 12 fields used in this
paper. Contours start at 10\% of the central value and increase by 10\%. As we
discuss below, this figure would be similar for other analytical functions
used.

\subsection{Orientation of the Galactic bar}

The characteristic angles of the fitted models ($\alpha,\beta$) are defined in
the same way for each model, so there is possible direct comparison between the
models. As can be seen from Table~2, the tilt angle $\beta$ is never important
in reducing the $\chi^2$ values significantly. Also, except for the G2 model,
angle $\beta$ is always smaller than $1\deg$. We will therefore set
$\beta=0\deg$ in further discussion.

\begin{figure}[t]
\plotfiddle{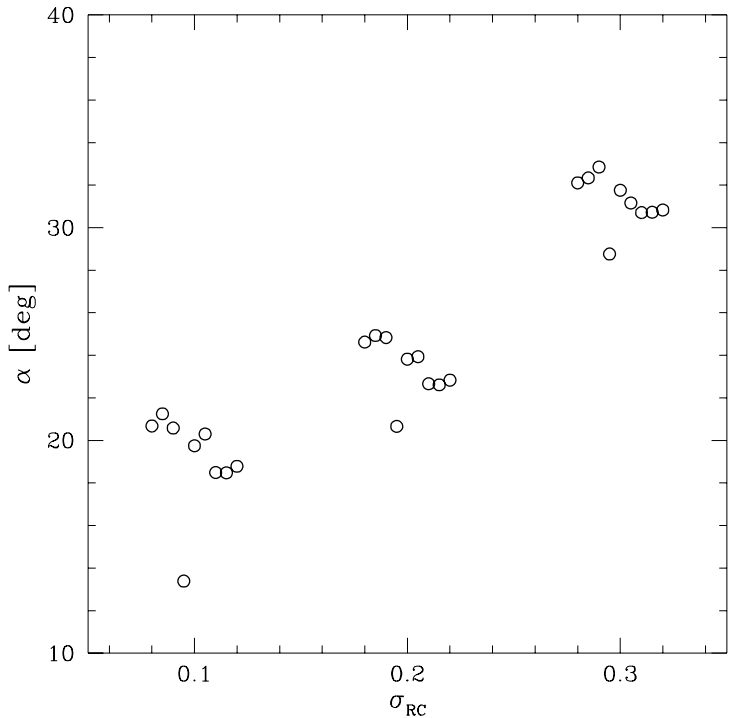}{8.5cm}{0}{50}{50}{-160}{-75}
\caption{The inclination angle for the various models at the assumed
width of the red clump $\sigma_{RC}=0.1,0.2,0.3\;mag$.  The horizontal spread
along the $x$-axis for each $\sigma_{RC}$ was done for the clarity.}
\label{fig7}
\end{figure}

To analyze the behavior of the angle $\alpha$ we plot in Fig.\ref{fig7} the
value of this angle for various models as the function of the assumed
$\sigma_{RC}$.  The horizontal spread along the $x$-axis for each $\sigma_{RC}$
was done for the clarity of presentation. It is immediately clear from
Fig.\ref{fig7} that for a given $\sigma_{RC}$ the values of $\alpha$ are very
similar between the models (see also Table~2). Even for the E1 model, which has
the most discrepant value of $\alpha$ comparing to other models, this value is
still only few degrees below the other values. Also, there is visible in
Fig.\ref{fig7} clear trend in the sense that for larger $\sigma_{RC}$ given
analytical model has larger inclination angle $\alpha$.  This trend was to some
extent expected: to fit the observed data, the models with larger $\sigma_{RC}$
have to have smaller depth along the line of sight, which for triaxial objects
decreases with increasing inclination angle.  For the preferred value of
$\sigma_{RC}=0.2\;mag$ (Section~2) the inclination angle is between
$\alpha\approx20-26\deg$. Even taking the most extreme values of $\alpha$ from
Fig.\ref{fig7} we find that the inclination angle is between
$\alpha\approx14-34\deg$. This is a very good constraint comparing to the
values of $\alpha$ found in the literature (for discussion see Dwek et
al. 1995; Gerhard 1996; Kuijken 1996).  Our result agrees roughly with Dwek's
et al.  range of $\alpha\approx0-40\deg$, with an average of
$\alpha\approx25\deg$.  In our case, however, constraints on the value of
$\alpha$ are much better due to the nature of the data used, which, unlike the
COBE-DIRBE data, contains the third dimension information in the form of the
shift between the red clump giants distributions (Fig.\ref{fig4}).

\subsection{Bar axis ratios}

\begin{figure}[t]
\plotfiddle{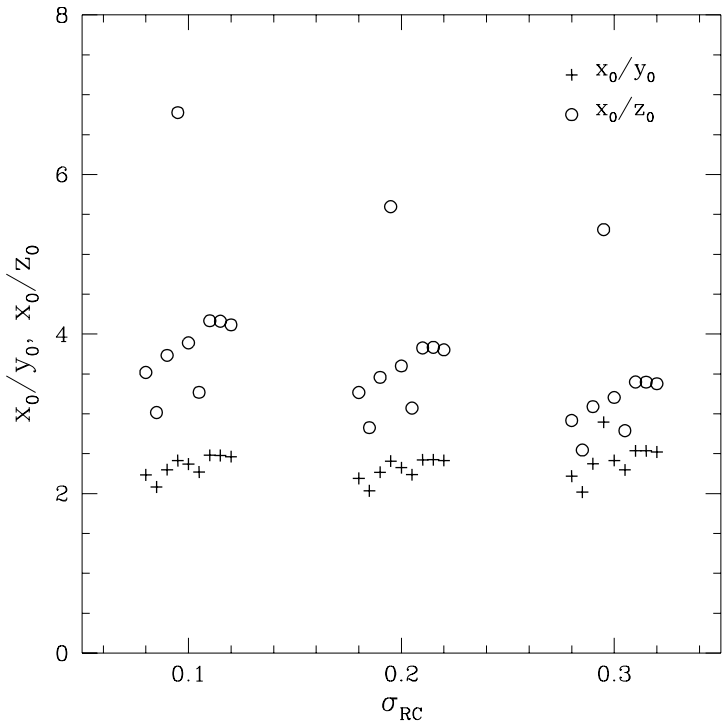}{8.5cm}{0}{50}{50}{-160}{-75}
\caption{The axis ratios $x_0/y_0,x_0/z_0$ for the various models at the
assumed width of the red clump $\sigma_{RC}=0.1,0.2,0.3\;mag$. The horizontal
spread along the $x$-axis for each $\sigma_{RC}$ was done for the clarity.}
\label{fig8}
\end{figure}

Unlike the characteristic angles, the scale lengths for different functional
forms cannot be compared directly. We can however compare the axis ratios for
various models. In Fig.\ref{fig8} we show the axis ratios $x_0/y_0,x_0/z_0$ for
the various models at the assumed width of the red clump
$\sigma_{RC}=0.1,0.2,0.3\;mag$. The horizontal spread along the $x$-axis for
each $\sigma_{RC}$ was done for the clarity of the presentation, as in
Fig.\ref{fig7}.

Let us first discuss the ratio $x_0/y_0$, i.e. the ratio of the bar major axis
to the minor axis lying in the plane of the Galaxy. This ratio for all the
models does not change significantly with $\sigma_{RC}$ and is between
$2.0-2.4$, except for the E1 model with $\sigma_{RC}=0.3\;mag$, for which
$x_0/y_0=2.9$. This is a very tight constraint, again much tighter than
$x_0/y_0\approx3\pm1$ obtained by Dwek et al. from COBE-DIRBE data (see also
their Fig.5).

On the other hand, there is a clear trend between $x_0/z_0$ and $\sigma_{RC}$
in the sense that for larger $\sigma_{RC}$ given analytical model has smaller
axis ratio $x_0/z_0$. This was somewhat expected: as we increase the
$\sigma_{RC}$, $x_0$ must decrease to fit the data, but we expect that $z_0$
would remain approximately the same, since $z_0$ is the scale length
perpendicular to the plane of the Galaxy and it is not strongly affected by
``rotating'' the bar.  Also, the scatter of $x_0/z_0$ for different models is
bigger than in case of $x_0/y_0$. For the preferred value of
$\sigma_{RC}=0.2\;mag$, $x_0/z_0$ is between $2.8-3.8$, again with the
exception of E1 model, for which it is $x_0/y_0\approx5.6$.

Judging from the map of the observed fields (Fig.\ref{fig3}), one could be
worried that most of the constraint we have on the $z_0$ comes from the TP8
field, in which there are very few red clump stars (Fig.\ref{fig4} and
Table~1). This is however not the case: even within Baade's Window there is a
clear gradient, in the number of the red clump stars, with the distance from
the Galactic center. The BW1 field has about 40\% more red clump stars than the
BW3 field. In fact, we run a series of fits without the TP8 field and the
resulting models were very close to what we present in this paper.

\subsection{Models with centroid offset}

There is evidence from the gas kinematics pointing to the fact that the
centroid of the gas emission from the Galactic bar is offset at $l\approx
0.5\deg$ (Blitz 1995; Weinberg 1996). We decided to add to our models one
additional parameter $\delta l$ which allows for the offset of the centroid of
the Galactic bar in the $l$ direction. In Table~3 we list the best fit
parameters of the resulting models, and also compare the $\chi^2/N_{DOF}$ for
models with the offset and without it.

\begin{planotable}{cccrrrcc}				       
\tablewidth{38pc}					       
\tablecaption{Parameters of the Best-Fit Bar Models with the Centroid Offset}
\tablehead{ \colhead{Model}   & \colhead{$\alpha$}   &	       
\colhead{$\delta l$} & \colhead{$x_0$}      & 		       
\colhead{$y_0$}   & \colhead{$z_0$}      &		       
 \colhead{$\chi^2/N_{DOF}$}  & \colhead{$\chi^2/N_{DOF}$} \\   
\colhead{}        & \colhead{[$\,\deg\,$]}     &	       
\colhead{[$\,\deg\,$]}  & \colhead{[$\,pc\,$]}      &		 
\colhead{[$\,pc\,$]}   & \colhead{[$\,pc\,$]}  & \colhead{\em with offset }
& \colhead{\em no offset}} 
\startdata							 
G1  & 24.1 & 0.60 & 1575 &  663 &  477 & 2.46 & 2.60 \nl
G2  & 24.1 & 0.82 & 1325 &  560 &  452 & 2.64 & 2.90 \nl
G3  & 23.9 & 0.53 & 4028 & 1656 & 1155 & 2.33 & 2.43 \nl
E1  & 19.1 & 0.24 & 1745 &  723 &  303 & 2.47 & 2.47 \nl
E2  & 23.5 & 0.46 &  935 &  385 &  262 & 2.29 & 2.37 \nl
E3  & 23.2 & 0.51 &  904 &  376 &  293 & 2.44 & 2.59 \nl
P1  & 22.5 & 0.18 &  784 &  321 &  207 & 2.26 & 2.27 \nl
P2  & 22.4 & 0.17 & 1095 &  448 &  288 & 2.26 & 2.27 \nl
P3  & 22.7 & 0.27 & 1486 &  606 &  396 & 2.25 & 2.27
\enddata
\end{planotable}

\begin{planotable}{ccrccc}				       
\tablewidth{30pc}					       
\tablecaption{Agreement of the Bar Models with the Artificial Data}
\tablehead{ \colhead{Model}   & \colhead{$\alpha$} &
\colhead{$x_0$}      & 		       
\colhead{$y_0/x_0$}   & \colhead{$z_0/x_0$}      &		       
 \colhead{$\chi^2/N_{DOF}$}  \\   
\colhead{}        & \colhead{[$\,\deg\,$]}     &	       
 \colhead{[$\,pc\,$]}      &		 
\colhead{}   & \colhead{}  
& \colhead{}} 
\startdata
E2  & 20.0 & 1000 & 1:2.00 & 1:2.50 & 1.11 \nl
\hline	          	       
G1  & 18.1 & 1478 & 1:1.96 & 1:2.42 & 1.19 \nl
G2  & 17.8 & 1298 & 1:1.87 & 1:2.27 & 1.28 \nl
G3  & 19.2 & 4537 & 1:1.97 & 1:2.63 & 1.09 \nl
E1  & 13.8 & 1650 & 1:1.90 & 1:3.48 & 1.40 \nl
E2  & 18.3 & 1000 & 1:1.98 & 1:2.58 & 1.07 \nl
E3  & 18.0 & 1075 & 1:1.91 & 1:2.43 & 1.12 \nl
P1  & 19.2 & 1565 & 1:2.00 & 1:2.69 & 1.14 \nl
P2  & 19.4 & 2653 & 1:2.01 & 1:2.72 & 1.16 \nl
P3  & 18.6 & 1869 & 1:1.99 & 1:2.62 & 1.08
\enddata
\end{planotable}

It is interesting to note that all the models prefer to have positive $\delta
l$, in agreement with Blitz~(1995).  Also, for some models, most noticeably the
G2 model, introducing this additional parameter reduced $\chi^2/N_{DOF}$
considerably. This is not simply an effect of having one additional parameter
-- the tilt angle $\beta$ had hardly any effect on our models. We conclude by
saying that our results are suggestive of a small ($\delta l\approx0.5\deg$)
centroid offset of the Galactic bar.

One could in principle introduce another additional parameter describing
the centroid offset in the $b$ direction. With current data, however,
we have only one field at positive Galactic latitude, which provides very
little lever arm to constrain an offset in $b$. We thus make no attempt
to include it in our models.

\section{FACTORS AFFECTING THE FITS}

In the previous section we fitted various analytical models to the luminosity
function of the bulge red clumps stars and obtained the parameters of the 3-D
mass distribution of the Galactic bar. This procedure is subject to possible
various systematic errors, which we discuss in some detail below.

\subsection{Bad selection of fitted models}

One obvious disadvantage of the whole procedure described in the previous
section is that we fit only a number of selected analytical functions
describing the 3-D light (mass) distribution of the Galactic bar. Indeed, high
value of the formal $\chi^2$ we obtained even for our best-fit model -- P1 with
$\sigma_{RC}=0.2\;mag$ and 8 free parameters has $\chi^2/N_{DOF}=2.27$ --
suggests that we either underestimate our errors (which were taken to be
Poissonian $\sqrt{N}$) or the Galactic bar 3-D mass distribution is described
by some other triaxial function or most probably is only roughly described by
an analytical function.  Nevertheless, by following Dwek et al. approach and
fitting a whole set of different analytical functions with different properties
we showed we can retrieve successfully at least some of the Galactic bar
properties.  For example, it is clear from Fig.\ref{fig7} that the inclination angle
of the bar to the line of sight $\alpha$ does not depend strongly on the model
we fit and is affected stronger by the assumed value of the intrinsic width of
the red clump $\sigma_{RC}$.

To investigate to some extent how strong the effect of fitting an improper
analytical function is on the fitted parameters, we generated artificial random
histograms of the $V_{_{V-I}}$ distributions using the E2 model.  We then
fitted this artificial data using all nine different analytical functions used
in this paper. In Table~4 we show both the parameters of the input model (first
row) and the fitted parameters of all nine models. For reasons discussed in the
previous section, instead of $y_0, z_0$ scale lengths we show the axis ratios
$y_0/x_0,z_0/x_0$, as these are more relevant for comparing the agreement
between the models.

As can be seen in Table~4, agreement in the fitted inclination angle is very
good. The E1 model deviates most from the input $\alpha=20\deg$, but this is
also true when fitting to the real data. This only reconfirms our strong
conviction that we can constrain the inclination angle $\alpha$ of the bar very
well. The $y_0/x_0$ axis ratio is also recovered remarkably well. As to the
$z_0/x_0$ axis ratio, except for the E1 model, it is also well recovered
by all other models. We conclude by saying that all the models we use,
except for the E1 model, can be used to recover the parameters of the
Galactic bar with similar fidelity. 

\subsection{Red clump detection efficiency}

It is possible that we detect red clump giants with different efficiency among
the 12 fields we use in this paper. This could affect some of the parameters we
fit, some stronger than others, for example inclination angle of the bar
probably would not depend very strongly on this effect, but the axis ratios
probably would. Red clump stars we use in our analysis are always well above
the detection threshold, $3-4\;mag$ in $V$, so we do not expect a strong
variation in efficiency among the fields.  Looking at fields which are
adjacent (like BW1 and BW5 or the two MM7 fields) we generally find that the
histograms of $V_{_{V-I}}$ values are similar for these fields (Fig.\ref{fig4} and
Table~1).

There are two levels on which stars can be lost: the CCD detection and the
construction of a CMD from $I$ and $V$ databases. Udalski et al.~(1993)
addressed the question of CCD star detection efficiency by placing artificial
stars on the CCD and then retrieving them through the standard OGLE data
pipeline. They found that the CCD detection efficiency in the interesting us
magnitude range is roughly constant and is about $\sim80\%$, which would not
change the parameters of the fit we are interested in. We therefore make no
correction for the CCD detection efficiency. The other reason that some stars
are not included in the CMD is that the star photometry is not good enough to
pass a corresponding criterion. Udalski et al.~(1993) used rather stringent
photometric quality constrains on what stars can qualify to be included in the
CMD, so only about 60\% of stars from the $I,V$ databases are finally
included. These constrains are however the same for all the fields, so we
neglect that in our fitting procedure.

\subsection{Effects of interstellar extinction}

One can imagine that a large scale gradient in the reddening law could explain
the systematic shift of the red clump giants in magnitude with the galactic
longitude, which we discuss in this and earlier works (Stanek et al.~1994;
1996). This would cause $V_{_{V-I}}$ defined by Eq.\ref{eq:free} to be invalid
globally. However, Wo\'zniak and Stanek~(1996) found no such gradient. What
they could not exclude were smaller variations of this coefficient among
various fields, but again, inspection of Fig.\ref{fig4} shows a good agreement
in red clump peak position for fields located at the approximately same
position, like the BW fields.  Even fields MM7 and GB13, located at
approximately the same galactic $l$, but on opposite sides of the Galactic
plane (Table~1), have practically the same red clump peak position
(Fig.\ref{fig4}), even that the GB13 field has somewhat higher extinction then
the MM7 fields.

\subsection{Galactic disk contamination}

As we select a region of the CMD and then use all stars in this region to
construct a luminosity function, we may expect some contamination from stars
located in the Galactic disk. Stanek et al.~(1994) considered this problem
using Bahcall \& Soneira~(1980) model of the Galactic disk and found the
contamination to be negligible (see Stanek's et al.~Fig.2). This is because in
the region corresponding to red clump giants we expect only either bright and
nearby, or evolved disk stars, of which there are very few.

A related problem: our star counts could also be contaminated by stars from the
metal poor stellar spheroid. We do not account for this population, having
already 8--9 free parameters in our fit. However, this should be negligible,
just as was contamination by disk stars. Red clump becomes weak quite rapidly,
as seen from $V_{_{V_I}}$ histogram of the fields GB11 and TP8, i.e. metal rich
Galactic bar star counts, in the region of the CMD we select for our analysis,
strongly dominate.

\subsection{Metallicity gradient in the Galactic bulge}

There seems to be a small metallicity gradient throughout the Galactic bulge
(Minniti et al.~1995; but see McWilliam \& Rich 1994), which could shift a
little intrinsic color and luminosity of the red clump stars across the
bulge. For all the models we took that into account by assuming
\begin{equation}
V_{_{V-I,RC}}=V_{_{V-I,0}}+\kappa\times\left\{
\begin{array}{l}
r \\
r_s,\\
r_e
\end{array}
\right.
\label{eq:colorshift}
\end{equation}
where the choice of $r,r_s$ or $r_e$ depends on the density model fitted.

We found that this additional parameter did not improve significantly the
quality of the fits, we therefore did not present in Table~2 the models with
$\kappa\neq0$.

\section{ASTROPHYSICAL IMPLICATIONS OF BEST-FIT MODELS}

\subsection{Agreement with COBE-DIRBE data}

The analytical models we used in this work were identical to those used by Dwek
et al., so we should be able to compare our results directly (with the
remainder that the angle $\alpha$ we use differs from Dwek et al.  $\alpha_D$
by $\alpha\equiv90-\alpha_D$). For comparison we use Dwek et al. best fitting
models at $2.2\;\micron$ with cutoff at $5\;kpc$, the same cutoff we use in
this paper (Eq.\ref{eq:integ}). The Dwek et al. fitting procedure (their
Table~1) does not like power-law models (our best fit models), so here we will
only compare six other models we both used.

\begin{figure}[t]
\plotfiddle{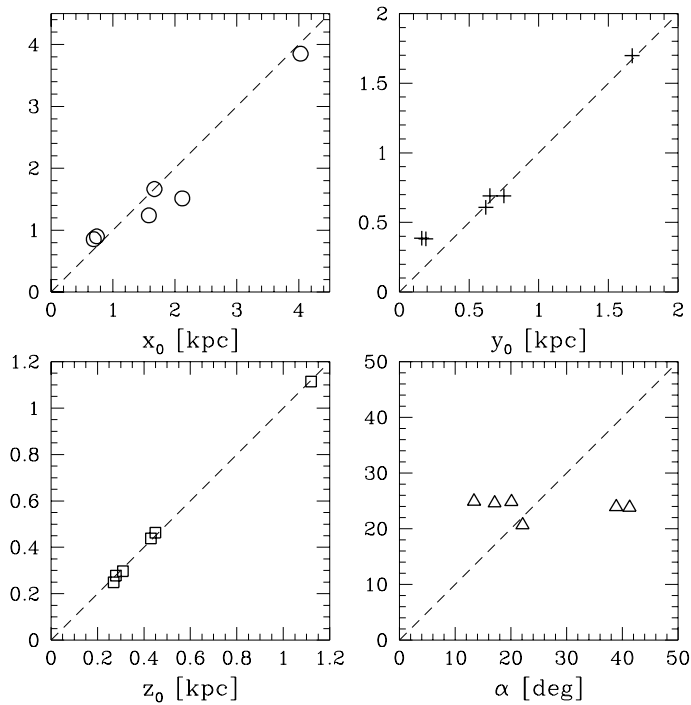}{8.5cm}{0}{50}{50}{-160}{-75}
\caption{The values of $x_0, y_0, z_0, \alpha$ from Dwek et al. ($x$-axis)
and from our work ($y$-axis). The dashed lines correspond to these values being
equal.}
\label{fig9}
\end{figure}

In Fig.\ref{fig9} we plot the values of $x_0, y_0, z_0, \alpha$ from Dwek et al.
($x$-axis) and from this work ($y$-axis). The dashed lines correspond to these
values being equal.  There is a very good agreement between the values of $z_0$
found by Dwek et al.  and those we find: they can be compared directly using
Dwek et al. Table~1 and our Table~2. This is extremely reassuring because $z_0$
is not affected by ``rotating'' the bar, so the fact that we both get
essentially the same values for $z_0$ means that, although here we are using
data of entirely different nature from COBE-DIRBE data, we probe in our
analysis the same stellar populations.

The correlations between $x_0, y_0$ values are not nearly as good as for $z_0$,
but still rather clear. On the other hand, there is no correlation between the
inclination angles $\alpha$ obtained by Dwek et al. and our values.  As we
discussed earlier in this paper, regardless of the fitted model we obtain
almost the same value for this angle, but Dwek et al.  obtained a wider range
of values for this angle. We are confident that the nature of the data we use
in this work allows for much better constraint of this parameter than the 2-D
COBE-DIRBE infrared intensity maps.

As we mentioned in Section~5, the G2 model which was Dwek et al. best fit model
fits our data worst. There is however several models which seem to fit both
COBE-DIRBE and our data reasonably well. In fact, for the G3 model the values
of the bar parameters obtained by Dwek et al.  (their Table~1) are remarkably
similar to ours (Table~2). The E2 model also fits reasonably well in both
cases, and we suspect that if the value of $\alpha$ for this model was fixed by
Dwek et al. to our value of $\alpha\approx24\deg$, the agreement between
$x_0,y_0$ would be much better. According to the trends we discussed in
Section~5, lowering $\alpha$ from Dwek et al. $41\deg$ to our $24\deg$ would
increase both $x_0$ and $y_0$ to values closer to ours. The E2 model might be
preferred over the G3 model as it fits better the COBE-DIRBE data in the inner
region of the Galaxy, where we do not have the CMD data from a low extinction
fields.

\begin{figure}[t]
\plotfiddle{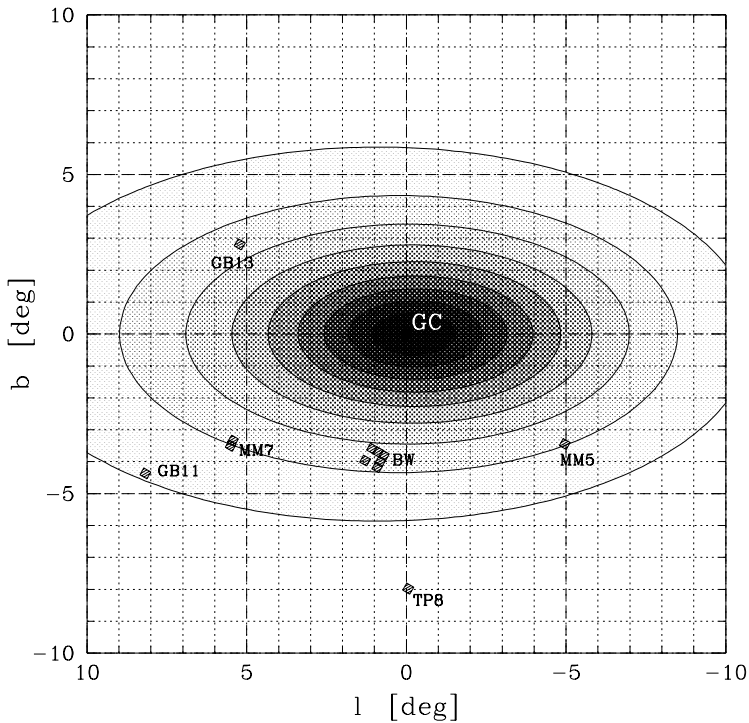}{8.5cm}{0}{50}{50}{-160}{-75}
\caption{Map of the projected light from the ``recommended'' E2 model in the
inner $20\times20\deg$ region of the Galaxy, with our fields also
shown. Contours start at 10\% of the central value and increase by 10\%.}
\label{fig10}
\end{figure}

To summarize, for those who want to use a Galactic bar model which should not
be grossly wrong over large range of galactocentric distances, we recommend the
E2 exponential model with {\em our\/} values for the bar parameters:
$\alpha=24\deg; x_0=900\;pc; y_0=385\;pc; z_0=250\;pc$.  In Fig.\ref{fig10} we
show the projected light from this model for the inner $20\times20\deg$ region
of the Galaxy, with our fields also shown. Contours start at 10\% of the
central value and increase by 10\%. The asymmetry between the positive and
negative $l$ sides of the Galactic bar is visible, but not striking.

\subsection{Microlensing phenomena due to the Galactic bar}

Microlensing can provide us with additional information on the properties of
the Galactic bar. As noticed first by Kiraga \& Paczy\'nski (1994), stars in
the Galactic bulge lensing other stars in the bulge produce significant
contribution to the microlensing optical depth, comparable to that produced by
the Galactic disk stars lensing bulge stars. This effect is strongly enhanced
when the Galactic bulge is a bar with the major axis inclined to the line of
sight by less the $45\deg$ (Paczy\'nski et al. 1994b; Zhao et al. 1995).
Recently Zhao \& Mao (1996) investigated in detail how different models of the
Galactic bar would differ in the optical depth for microlensing they produce,
so here we want to show only some of the aspects of the microlensing by the
Galactic bar.

\begin{figure}[t]
\plotfiddle{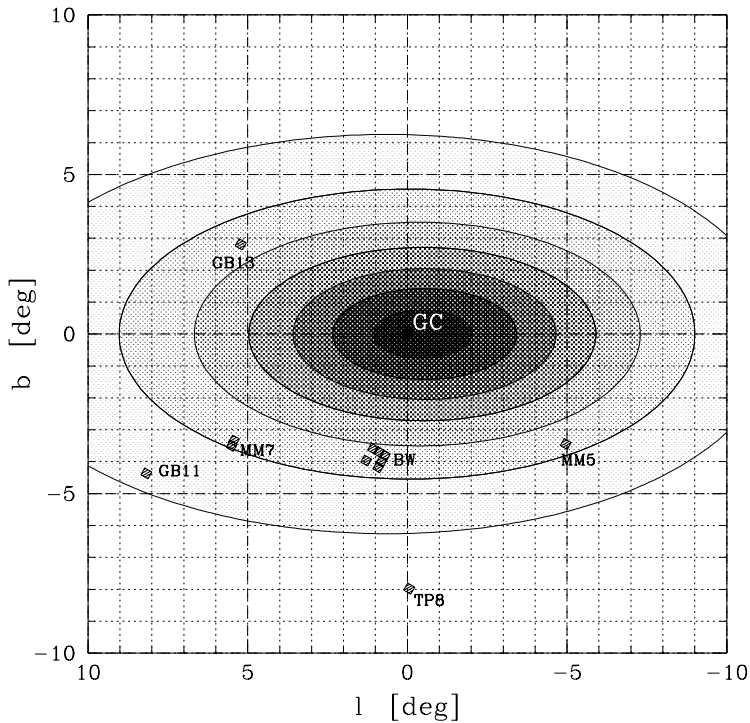}{8.5cm}{0}{50}{50}{-160}{-75}
\caption{Map of the microlensing optical depth $\tau$ produced by the
``recommended'' E2 model in the inner $20\times20\deg$ region of the Galaxy,
with our fields also shown. The mass of the Galactic bar was assumed to be
$M_{bar}=2\times10^{10}\;m_{\odot}$.  Contours start at $\tau=5\times10^{-7}$
and increase also by $\tau=5\times10^{-7}$ up to $3.5\times10^{-6}$. The
contribution to the optical depth from the Galactic disk was not included.}
\label{fig11}
\end{figure}

In Fig.\ref{fig11} we show a map of the microlensing optical depth $\tau$
produced by our ``recommended'' E2 model in the inner $20\times20\deg$ region
of the Galaxy, with our fields also shown. The mass of the Galactic bar was
assumed to be $M_{bar}=2\times10^{10}\;m_{\odot}$.  Contours start at
$\tau=5\times10^{-7}$ and increase also by $\tau=5\times10^{-7}$ to the maximum
of $\tau=3.5\times10^{-6}$. The contribution to the optical depth from the
Galactic disk was not included.  Details of this microlensing map would be
different if we used one of other models discussed in this paper, but the
overall effect of the optical depth dropping off rapidly with the distance from
the Galactic center, faster along the minor axis, would remain. As discussed by
Zhao \& Mao (1996), some few hundred microlensing events are needed before the
microlensing can really make the difference for the existing Galactic bar
models. Fortunately, this is very feasible judging from the existing
microlensing experiments observing the Galactic bar (Udalski et al. 1994;
Alcock et al. 1995; Alard et al. 1995; Alcock et al. 1996).

Another aspect of the microlensing by the Galactic bar was first noticed by
Stanek (1995). As can be deduced from Fig.\ref{fig6}, the stars from the far
side of the Galactic bar will be preferentially lensed if most of the lensing
comes from the bar stars (Paczy\'nski et al. 1994b; Zhao et al. 1995). This
leads to the offset in the brightness of the lensed stars, as compared to all
observed stars, discussed in detail by Stanek (1995). The value of this offset
correlates very well with the depth of the Galactic bar along the line of sight
(Stanek 1995, his Fig.3). About a hundred of microlensing events in the red
clump region of the CMD are needed to measure this effect with desired
accuracy.

\section{SUMMARY}

In this paper we have used color-magnitude data obtained by the OGLE
collaboration for 12 fields scattered across the galactic bulge
(Fig.\ref{fig3}) to construct the three-dimensional model of the mass
distribution in the Galactic bar.  We model the Galactic bar by fitting for all
fields the observed luminosity functions in the red clump region of the
color-magnitude diagram (Fig.\ref{fig4}, \ref{fig5}).  We find that almost
regardless of the analytical function used to describe the 3-D stars
distribution of the Galactic bar, the resulting models are inclined to the line
of sight by $20-30\deg$ (Fig.\ref{fig7}), with axis ratios corresponding to
$x_0\!:\!y_0\!:\!z_0=3.5\!:\!1.5\!:\!1$ (Fig.\ref{fig8}). This puts a strong
constraint on the possible range of the Galactic bar models.

Comparing our result with the results of Dwek et al. (1995), we find a good
agreement between the derived parameters of the models (Fig.\ref{fig9}), but we
constrain the bar parameters much tighter. We recommend as a model which fits
well both COBE-DIRBE and our data the E2 model (Eq.9) with the values for the
bar parameters: $\alpha=24\deg; x_0=900\;pc; y_0=385\;pc; z_0=250\;pc$.  We
show various properties of this model in Fig.ref{fig6}, \ref{fig10},
\ref{fig11}.  Gravitational microlensing can provide us with additional
constrains on the structure of the Galactic bar.

The text of this paper along with the figures in PostScript format
is available using anonymous {\tt ftp} on {\tt astro.princeton.edu},
in {\tt stanek/Barmodel} directory.

\acknowledgments{We would like to thank B. Paczy\'nski for constant advice and
many stimulating discussions.  W. Colley once again helped us to ``polish'' the
text. KZS was supported by the NSF grant AST 9216494 to Bohdan Paczy\'nski and
also with the NAS Grant-in-Aid of Research through Sigma Xi, The Scientific
Research Society.  AU, MSz and MK were supported with Polish KBN grant
2P03D02908 and JK was supported with Polish KBN grant 2P03D00808.}

\end{document}